\documentclass[twocolumn]{aastex631}
\usepackage{amsmath}

\newcommand{\pa}{\partial}

\usepackage[T1]{fontenc}

\shorttitle{The Gamma Ray Emission from RX~J1713.7-3946}
\shortauthors{Fujita et al.}

\begin{document}

\title{The Gamma-Ray Emission from the Supernova
Remnant RX~J1713.7-3946 Interacting with Two-phase Medium}

\author[0000-0003-0058-9719]{Yutaka Fujita}
\affiliation{Department of Physics, Graduate School of Science, 
Tokyo Metropolitan University,\\
1-1 Minami-Osawa, Hachioji-shi, Tokyo 192-0397, Japan}

\author[0000-0002-1251-7889]{Ryo Yamazaki}
\affiliation{Department of Physical Sciences, Aoyama Gakuin University, 
5-10-1
Fuchinobe, Sagamihara 252-5258, Japan}
\affiliation{Institute of Laser Engineering, Osaka University, 
2-6 Yamadaoka,
Suita, Osaka 565-0871, Japan}

\author[0000-0002-2387-0151]{Yutaka Ohira}
\affiliation{Department of Earth and Planetary Science, 
The University of Tokyo, 7-3-1 Hongo, Bunkyo-ku,
Tokyo 113-0033, Japan}

\begin{abstract}

We study the origin of gamma rays from the supernova remnant (SNR)
RX J1713.7-3946. Using an analytical model, we calculate the
distribution of cosmic rays (CRs) around the SNRs. Motivated by the
results of previous studies, we assume that the SNR is interacting with
two-phase interstellar medium (ISM), where dense clumps are surrounded
by tenuous interclump medium. We also assume that only higher-energy
protons ($\gtrsim$~TeV) can penetrate the dense clumps. We find that
$\pi^0$-decay gamma rays produced by protons reproduce the observed
gamma-ray spectrum peaked at $\sim$TeV. On the other hand, it
has recently been indicated that the observed ISM column density
($N_p$), the X-ray surface brightness ($I_X$), and the gamma-ray surface
brightness ($I_g$) at grid points across the SNR form a plane in the
three-dimensional (3D) space of ($N_p$, $I_X$, $I_g$). We find that
the planar configuration is naturally reproduced if the ISM or
the CR electron-to-proton ratio is not spherically uniform. We show that
the shift of the observed data in the 3D space could be used to identify
which of the quantities, the ISM density, the CR electron-to-proton
ratio, or the magnetic field, varies in the azimuthal direction of the
SNR.

\end{abstract}

\keywords{Supernova remnants (1667) --- Interstellar medium (847) ---
Cosmic ray sources (328) --- Gamma-ray sources (633) --- X-ray sources
(1822)}

\section{Introduction} \label{sec:intro}

Supernova remnants (SNRs) are the most promising accelerators of
cosmic rays (CRs) below the knee ($\sim 3\times 10^{15}$~eV). CRs
are believed to be accelerated through a diffusive shock-acceleration
mechanism (DSA;
\citealt{1978ApJ...221L..29B,1978MNRAS.182..147B,1987PhR...154....1B}).
Nonthermal emissions from the accelerated CRs have been observed. The
detections of synchrotron X-rays in some SNRs have been regarded as
evidence for the acceleration of electrons to ultrarelativistic energies
at SNR shocks \citep{1995Natur.378..255K}. On the other hand, it is
still uncertain whether the gamma rays from SNRs are generated through
either leptonic (inverse Compton, IC, scattering of low-energy photons
by high-energy electrons) or hadronic ($\pi^0$-decay photons
generated through $pp$-interaction) processes. The sharp cutoff at low
energies ($\sim 100$~MeV) in the gamma-ray spectra of the middle-aged
SNRs IC~443 and W44 can be regarded as the so-called $\pi^0$-bump,
which is direct proof for the hadronic origin of the gamma rays
\citep{2011ApJ...742L..30G,2013Sci...339..807A}. On the other hand, for
the young SNR RX J1713.7-3946 (RX~J1713 hereafter),
\citet{2010ApJ...712..287E} indicated that the hadronic model of
gamma-ray emission cannot reproduce the observed X-ray emission because of
an overproduction of thermal X-ray line emission, using 1D
hydrodynamic simulations of a supernova blast wave. However, it has
been claimed that the thermal X-ray line emission can be faint if the
SNR is interacting with inhomogeneous interstellar medium 
\citep[ISM;][]{2010ApJ...708..965Z,2012ApJ...744...71I,2014MNRAS.445L..70G}. These
studies also indicated that the spectrum of the $\pi^0$-decay gamma rays
generated by CR protons can mimic that of leptonic gamma rays if the ISM
is inhomogeneous.

In general, angular resolutions of gamma-ray detectors are worse than
those of radio and X-ray telescopes. Thus, gamma-ray observations with
high angular resolutions could change the situation. Recently,
\citet{2021ApJ...915...84F} analyzed H.E.S.S gamma-ray data of RX~J1713
with improved angular resolutions. They assumed that the gamma-ray
counts are given as a linear combination of two terms: one is
proportional to the ISM column density, and the other proportional to the
X-ray counts. By fitting the expression to the data pixels, 
that is to say, the data at grid points across the SNR, they discovered
that the gamma-ray counts are well represented by a plane in 3D space formed by the ISM column density, the X-ray
counts, and the gamma-ray counts (see Figure~4 in
\citealt{2021ApJ...915...84F} or Figure~\ref{fig:plane_nav} in this
paper). They indicated that the plane angle suggests that the hadronic
and leptonic components constitute (67 $\pm$ 8)\% and (33 $\pm$ 8)\% of
the total gamma rays, respectively.

In this paper, we study gamma ray emissions from an SNR interacting with
inhomogeneous ISM taking RX~J1713 as an example. We also discuss the
implications of the plane discovered by \citet{2021ApJ...915...84F}. In
particular, we take account of the fact that an SNR is rather
spherically symmetric, which was not explicitly considered by
\citet{2021ApJ...915...84F}. This paper is organized as follows. In
Section~\ref{sec:model}, we describe our models for the CR distributions
and the ISM.  In Section~\ref{sec:result}, we show the results of our
model and discuss the spectral energy distribution and the plane. The
conclusion of this paper is presented in Section~\ref{sec:conc}.

\section{Models}
\label{sec:model}

\subsection{Cosmic-ray Distributions}
\label{sec:CRdist}

\subsubsection{Outside the Supernova Remnant}

We derive the distribution function of CR protons outside the
SNR based on the model by \citet{2011MNRAS.410.1577O}. It can be
obtained by solving a diffusion equation,
\begin{equation}
\label{eq:diff}
\frac{\partial n_p}{\partial t}(t,\mbox{\boldmath $r$},p) 
- D_{\rm ISM}(p) \Delta n_p(t,\mbox{\boldmath $r$},p) 
= q_{s}(t,\mbox{\boldmath $r$},p)\:,
\end{equation}
where $\mbox{\boldmath $r$}$ is the position, $p$ is the CR momentum,
$n_p$ is the distribution function, $D_{\rm
ISM}(p)$ is the diffusion coefficient for CRs, and $q_{\rm s} (t,\mbox{\boldmath $r$},p)$ is the source
term of CRs. We assume that the SNR is spherically symmetric and
$r$ is the distance from the SNR center. It is also assumed that CRs
with a momentum $p$ escape from the SNR at $t=t_{\rm esc}(p)$
\citep{2005A&A...429..755P,2010A&A...513A..17O}. For a point
source, the source term is written as $q_{\rm s} = N_{\rm
esc}(p)\delta(\mbox{\boldmath $r$} )\delta[t-t_{\rm esc}(p)]$, and the
solution is
\begin{equation}
\label{eq:fpoint}
 N_{\rm point}(t,r,p) 
= \frac{\exp[-(r/R_{\rm d})^2]}{\pi^{3/2}R_{\rm d}^3}N_{\rm esc}(p)\:,
\end{equation}
where
\begin{equation}
 R_{\rm d}(t,p) = \sqrt{4\: D_{\rm ISM}(p)[t-t_{\rm esc}(p)]}\:,
\end{equation}
and 
\begin{equation}
 N_{\rm esc}(p)=\int dt\int d^3\mbox{\boldmath $r$}\: q_{\rm s} (t,r,p)
\:,
\end{equation}
which represents the spectrum of the whole CRs.

In the case of a spherical SNR, CRs escape from a surface, $R_{\rm
esc}(p)$. Thus, the source term is given by
\begin{equation}
\label{eq:qsSNR}
q_s(p) = \frac{N_{{\rm esc}}(p)}{4\pi r^2} 
\delta[r-R_{{\rm esc}}(p)]\delta[t-t_{{\rm esc}}(p)]\:.
\end{equation}
The solution of equation~(\ref{eq:diff}) can be obtained using
equation~(\ref{eq:fpoint}) as the Green function,
\begin{eqnarray}
\label{eq:Np_out}
n_p(t,r,p) &=& \int d^3\mbox{\boldmath $r$}' N_{\rm point}(t,|r-r'|,p) 
\frac{\delta[r'-R_{\rm esc}(p)]}{4\pi r'^2}\nonumber\\ 
 &=& \frac{ e^{-(\frac{r-R_{{\rm esc}}(p)}{R_{\rm d}(t,p)})^2 } 
- e^{-(\frac{r+R_{{\rm esc}}(p)}{R_{\rm d}(t,p)})^2 } }
{4\pi^{3/2}R_{\rm d}(t,p)R_{{\rm esc}}(p)r} N_{{\rm esc}}(p)\:.
\end{eqnarray}

We need to specify $R_{\rm esc}(p)$, $t_{\rm esc}(p)$, and $N_{\rm
esc}(p)$. In this study, we consider the case where the SNR is in the
Sedov phase at least until recently. Thus, the shock radius is
represented by
\begin{equation}
\label{eq:Rsh}
 R_{\rm sh}(t) 
= R_{\rm Sedov}\left(\frac{t}{t_{\rm Sedov}}\right)^{2/5}\:,
\end{equation}
and the escaping radius is given by
\begin{equation}
\label{eq:Resc}
 R_{\rm esc}(t) = (1+\kappa)R_{\rm sh}(t)\:, 
\end{equation}
where we adopt $\kappa=0.04$
\citep{2005A&A...429..755P,2010A&A...513A..17O}.  As long as $\kappa\ll
1$, the difference between $R_{\rm esc}$ and $R_{\rm sh}$ can be ignored
in the following discussion. We assume that CRs with $p=p_{\rm esc}$
escape from the surface ($r=R_{\rm esc}$) at $t=t_{\rm esc}$. We expect
that the escape momentum $p_{\rm esc}$ is a decreasing function of the
shock radius. Thus, we adopt a phenomenological power-law relation,
\begin{equation}
\label{eq:pesc}
p_{\rm esc} = p_{\rm max}
\left(\frac{ R_{\rm sh} }{ R_{\rm Sedov} } \right)^{-\alpha}\:,
\end{equation}
where $p_{\rm max}$ and $R_{\rm Sedov}$ are the escape momentum and the
shock radius at the beginning of the Sedov phase ($t=t_{\rm Sedov}$),
respectively. We set $R_{\rm Sedov}=2.1$~pc, $t_{\rm Sedov}=210$~yr, and
$\alpha=6.5$ following \citet{2011MNRAS.410.1577O}, although these
values are not particularly adjusted to RX~J1713. The maximum momentum
$p_{\rm max}$ is a parameter, and we fix this at $p_{\rm
max}c=10^{15}\rm\: eV$ to reproduce the observed gamma-ray spectrum.
Eliminating $R_{\rm sh}$ from equations~(\ref{eq:Rsh}) and
(\ref{eq:pesc}), and replacing $p_{\rm esc}$ and $t$ with $p$ and
$t_{\rm esc}$, respectively, we obtain
\begin{equation}
\label{eq:tesc}
 t_{\rm esc}(p) = t_{\rm Sedov}\left(\frac{p}{p_{\rm max}}\right)
^{-5/(2\alpha)}\:.
\end{equation}

We assume that the CR spectrum at the shock front is given by 
\begin{equation}
\label{eq:npRsh}
 n_p(p,R_{\rm sh}) = A \left(\frac{p}{p_{\rm max}}\right)^{-s} 
\left(\frac{R_{\rm sh}}{R_{\rm Sedov}}\right)^{\beta-3}\:,
\end{equation}
where $A$ is the normalization. In the following, we adopt $s=2$ and
$\beta=3(3-s)/2$; the latter is the relation obtained for a thermal
leakage model for CR injection
\citep{2010A&A...513A..17O,2011MNRAS.410.1577O}. The spectrum of the
escaped CRs ($p>p_{\rm esc}$) can be written as
\begin{equation}
\label{eq:Nesc}
 N_{\rm esc}(p)\propto p^{-(s+\beta/\alpha)}\:,
\end{equation}
\citep{2010A&A...513A..17O}. The normalization is determined from the
total energy of the CR protons with $pc>1$~TeV ($E_{\rm tot,CR}$), which
is treated as a parameter.

Outside the SNR ($r>R_{\rm esc}\sim R_{\rm sh}$), the diffusion
coefficient is assumed to be
\begin{equation}
\label{eq:DISM}
D_{\rm ISM}(p) = 10^{28}\:\chi \left(\frac{pc}{10\:\rm GeV}\right)^\delta
\rm cm^{2} s^{-1}\:
\end{equation}
\citep{2011MNRAS.410.1577O}. We adopt Kolmogorov-type turbulence
($\delta = 1/3$), which is theoretically motivated
\citep[see also][]{2021ApJ...909...46T} and close to those derived from
observations ($\delta \sim 0.4$;
\citealt{2015JCAP...12..039E,2015A&A...580A...9G}). The constant $\chi
(\leq 1)$ is introduced because the coefficient around SNRs can be
significantly reduced by waves created through the streaming of escaping
CRs
(e.g. \citealt{1975MNRAS.173..255S,2010ApJ...712L.153F,2011MNRAS.415.3434F}). In
this study, we fix it at $\chi=0.01$ \citep{2009ApJ...707L.179F}.

In this way, equation~(\ref{eq:Np_out}) is determined. We note that the
equation is applied to escaped CRs ($p>p_{\rm esc}$). It is assumed that
the functional form of CR electron density $n_e(t,r,p)$ is also
represented by equation~(\ref{eq:Np_out}) if radiative cooling is
inefficient. The ratio to $n_p$ is assumed to be $K_{\rm ep}=0.05$ in the
fiducial model, which is consistent with the observed spectral energy
distribution (SED; see Figure~\ref{fig:SED_fid}).  In contrast with
protons, however, electrons suffer from radiative cooling such as
synchrotron emission, IC scattering, and nonthermal
bremsstrahlung. Thus, if the cooling time of the electrons $t_{{\rm
cool},e}$ is longer than $t-t_{\rm esc}(p)$, the density is
$n_e(t,r,p)=0$.

\subsubsection{Inside the Supernova Remnant}
\label{sec:inside}

For the sake of simplicity, we assume that the diffusion coefficient for
CRs is almost zero at $r<R_{\rm sh}$ because of efficient scattering of
CR protons by turbulence developed down the shock. In this case, the
distribution of CRs depends on advection and adiabatic loss and can be
obtained by solving the following equation
\citep{2005A&A...429..755P,2010A&A...513A..17O}:
\begin{equation}
\label{eq:adv}
 \frac{\pa n_p}{\pa t} + u \frac{\pa n_p}{\pa r} 
- \frac{p^2}{3r^2}\frac{\pa}{\pa r}(r^2 u)p\frac{\pa }{\pa p}\left(\frac{n_p}{p^2}\right) = 0\:,
\end{equation}
where $u$ is the gas velocity, and we simply set
\begin{equation}
u(r,t) = \left(1-\frac{1}{\sigma}\right)
\frac{r}{R_{\rm sh}(t)}u_{\rm sh}(t)\:,
\end{equation}
where $\sigma$ is the compression ratio of the shock and $u_{\rm sh}$ is
the shock velocity \citep{2005A&A...429..755P}. In the following, we
adopt $\sigma=4$.

Given equation~(\ref{eq:npRsh}), the solution of equation~(\ref{eq:adv})
is written as
\begin{eqnarray}
\label{eq:Np_in}
 n_p(t,r,p) &=& A \left(\frac{p}{p_{\rm max}}\right)^{-s} 
\left(\frac{r}{R_{\rm Sedov}}\right)^{\sigma(s+\beta-1)-s-2}\nonumber \\ 
& &\times
\left(\frac{R_{\rm sh}(t)}{R_{\rm Sedov}}\right)^{-(s+\beta-1)(\sigma-1)}\:.
\end{eqnarray}
In equation~(\ref{eq:Np_in}), the normalization $A$
can be determined from $N_{\rm esc}(p)$ (equation~(\ref{eq:Nesc})),
because it is written as
\begin{equation}
\label{eq:Nesc_in}
 N_{\rm esc} = N_{\rm esc,surface} + N_{\rm esc,inside}\:,
\end{equation}
where first term on the right-hand side describes the particles that run
away from the shock front, and the second term describes the particles
escaping from the shock interior \citep{2005A&A...429..755P}.  With the
help of equations~(\ref{eq:pesc}) and (\ref{eq:npRsh}), they are
represented as
\begin{equation}
\label{eq:Nesc,surface}
 N_{\rm esc,surface}(p) = \frac{4\pi A}{\alpha}
\frac{1-\frac{1}{\sigma}-\frac{\xi_{\rm cr}}{2}}{3}
\left(\frac{p}{p_{\rm max}}\right)^{-s-\beta/\alpha}\:,
\end{equation}
and
\begin{eqnarray}
\label{eq:Nesc,inside}
 N_{\rm esc,inside}(p) &=& \frac{4\pi A}{\alpha}
\frac{\sigma-1-\frac{1}{\sigma}}{1-s-\sigma(s+\beta-1)}\nonumber \\
& &\times \left(\frac{p}{p_{\rm max}}\right)^{-s-\beta/\alpha}\:,
\end{eqnarray}
respectively \citep{2010A&A...513A..17O}. In
equation~(\ref{eq:Nesc,surface}), $\xi_{\rm cr}$ is the ratio of the CR
pressure to the shock ram pressure, and we adopt $\xi_{\rm cr}=0.5$
\citep{2005A&A...429..755P}.  We note that
Equation~(\ref{eq:Nesc,inside}) is realized when
$\beta>(s-1)(1/\sigma-1)$, which is consistent with our chosen
parameters. Thus, once the normalization of equation~(\ref{eq:Nesc}) is
given, normalization $A$ in equation~(\ref{eq:Np_in}) is derived
from equations~(\ref{eq:Nesc_in})--(\ref{eq:Nesc,inside}).

We note that equation~(\ref{eq:Np_in}) is applied for CR protons that
have not escaped from the SNR ($p<p_{\rm esc}$). Thus, $n_p=0$ for
$p>p_{\rm esc}$ at $r<R_{\rm sh}$ because they have escaped. Moreover,
considering that the advection time of CRs,
\begin{equation}
 t_{\rm adv}\sim\frac{R_{\rm sh}(t)-r}{u_{\rm sh}/\sigma}\:,
\end{equation}
is finite, we also assume that $n_p=0$ at the radius $r$ that satisfies
$t<t_{\rm adv}$. The distribution of CR electrons is given by
$n_e=K_{\rm ep}n_p$, where $K_{\rm ep}$ is the electron-to-proton ratio, if
cooling is ignored.

\subsection{Interstellar Medium}

\begin{figure}
\includegraphics[width=84mm]{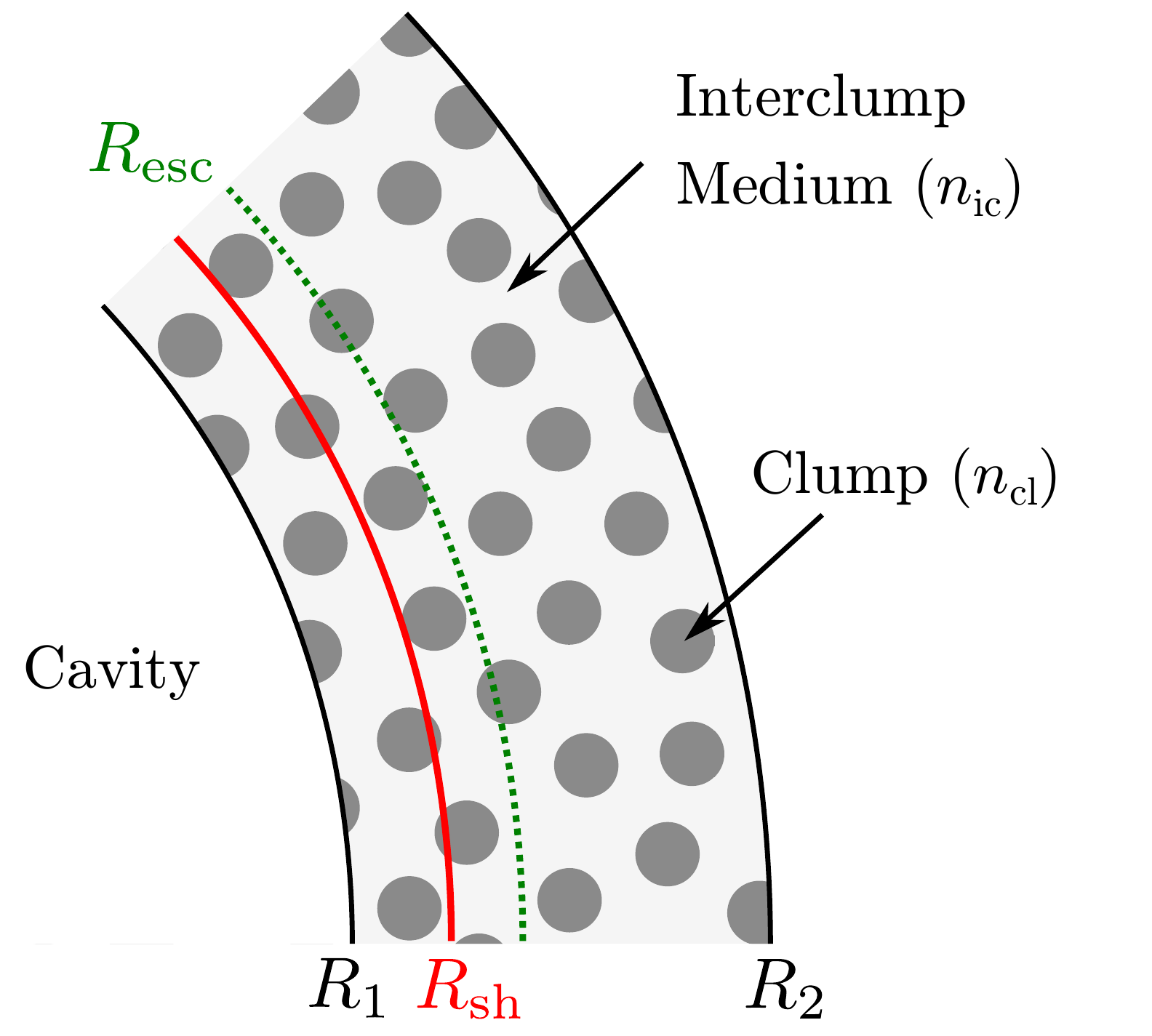} \caption{Schematic of the
SNR. The ISM surrounding the cavity consists of dense clumps with
density $n_{\rm cl}$ and interclump medium with density $n_{\rm
ic}$. The inner and outer boundaries are represented by $r=R_1$ and
$R_2$, respectively. The SNR only recently hit the inner boundary, and the
shock radius ($R_{\rm sh}$) and the escape radius ($R_{\rm esc}$) are
located inside the ISM}  \label{fig:ISM}
\end{figure}

In our model, the supernova explosion occurred inside a low-density
cavity produced by stellar winds driven by the progenitor massive star.
We also assume that the SNR only recently hit the dense ISM surrounding
the cavity, which we simply refer to as ``the ISM'' from now
on. Motivated by the results of numerical simulations
\citep{2012ApJ...744...71I}, we consider a two-phase medium, in which
dense clumps are immersed in the tenuous interclump medium
(Figure~\ref{fig:ISM}). Thus, we assume that the density of the
interclump medium ($n_{\rm ic}$) is much lower than that of the clumps
($n_{\rm cl}$), and the volume filling factor of the clumps is very
small ($x\ll 1$). This extreme density contrast is required to
avoid strong thermal emission from the interclump medium (see
Section~\ref{sec:SED}), and it can be realized if stellar winds from the
progenitor star have blown out the low-density ISM before the explosion
of the supernova (see Figure~9 in \citealt{2012ApJ...744...71I}). The
average density of the ISM is represented by
\begin{equation}
\label{eq:nav}
 n_{\rm av}=x n_{\rm cl} + (1-x)n_{\rm ic}\:.
\end{equation}

The inner and outer boundaries of the ISM are located at $R_1=6$~pc and
$R_2=9$~pc, respectively (Figure~\ref{fig:ISM}). We assume that
the stellar winds had blow away clumps at $r<R_1$; this is required so
that the profile of $N_p$ is consistent with observations (see
Figure~\ref{fig:profile}). For $R_{\rm sh} < r < R_2$ (upstream), the
average density of the ISM is $n_{\rm av}=250\rm\: cm^{-3}$ and the
density of the interclump medium is $n_{\rm ic}=0.025\rm\: cm^{-3}$ in
the fiducial model. Assuming that the filling factor is $x=0.01$, the
density of the clumps is $n_{\rm cl}\approx 2.5\times 10^4\rm\: cm^{-3}$
(equation~(\ref{eq:nav})). We choose these values to approximately
reproduce observational results for the ISM
\citep[e.g.][]{2015ApJ...799..175S,2021ApJ...915...84F}. The
density contrast between the clumps and the interclump medium is fairly
high ($n_{\rm cl}/n_{\rm ic}\approx 10^6$). The magnetic fields that
are compressed during the formation of the clumps are likely to support
the clumps.  It is assumed that $n_{\rm ic}$ at $R_1 < r < R_{\rm sh}$
(downstream) is $\sigma~(=4)$ times larger than that of the upstream
value, which means $n_{\rm ic}=0.1\rm\: cm^{-3}$.\footnote{This
low value of $n_{\rm ic}$ may suggest that the mass of the supernova
ejecta is not too high (e.g. lower than a few $M_\odot$.). Or part of
the ejecta is not observed as diffuse X-ray gas because it is highly
clumpy and/or has not been heated by the reverse shock
\citep{2010ApJ...708..965Z}.} This is consistent with the value obtained
by \citet{2015ApJ...814...29K}, and it does not affect the results as
long as $n_{\rm ic}\ll n_{\rm av}$. On the other hand, we assume that
$n_{\rm cl}$ at the downstream side is the same as that at the upstream
side because numerical simulations showed that the density of the
clumps is not much affected by the passage of the shock
\citep{2012ApJ...744...71I}. Thus, the average density $n_{\rm av}$ at
the downstream side is almost same as that at the upstream side as long
as $x\ll 1$ and $x\sim\rm const$ (equation~(\ref{eq:nav})). The
size of individual clumps does not appear in our formulation. Numerical
simulations showed that when a shock front passes a clump, the clump
maintains its density contrast and the front is not much distorted
(\citealt{2019MNRAS.487.3199C}, see also
\citealt{2017ApJ...846...77S,2018MNRAS.478.2948W}). This suggests that
the clump size does not significantly influence the clump sustainability
and the shock propagation, although clumps that are too small may be destroyed at
the shock passage.

The magnetic fields in the ISM ($B$) affect synchrotron emissions from
CR electrons. We assume that the effective magnetic field
strength in the ISM is $B_u=12\:\mu\rm G$ at the upstream side
($r>R_{\rm sh}$) and $B_d=\sigma B_u = 48\:\mu\rm G$ at the downstream
side ($r<R_{\rm sh}$). We do not mean that the magnetic
fields in the interclump space are uniform; they can be intensified
locally. In fact, numerical simulations showed that they are amplified
around the clumps due to turbulence that develops through the passage of the
shock (\citealt{2012ApJ...744...71I}, see also
\citealt{2007Natur.449..576U}). Synchrotron emissions from inside the
clumps can be ignored because of the small filling factor.

\section{Results}
\label{sec:result}

\subsection{Spectral Energy Distribution}
\label{sec:SED}

\begin{figure}[t]
\includegraphics[width=84mm]{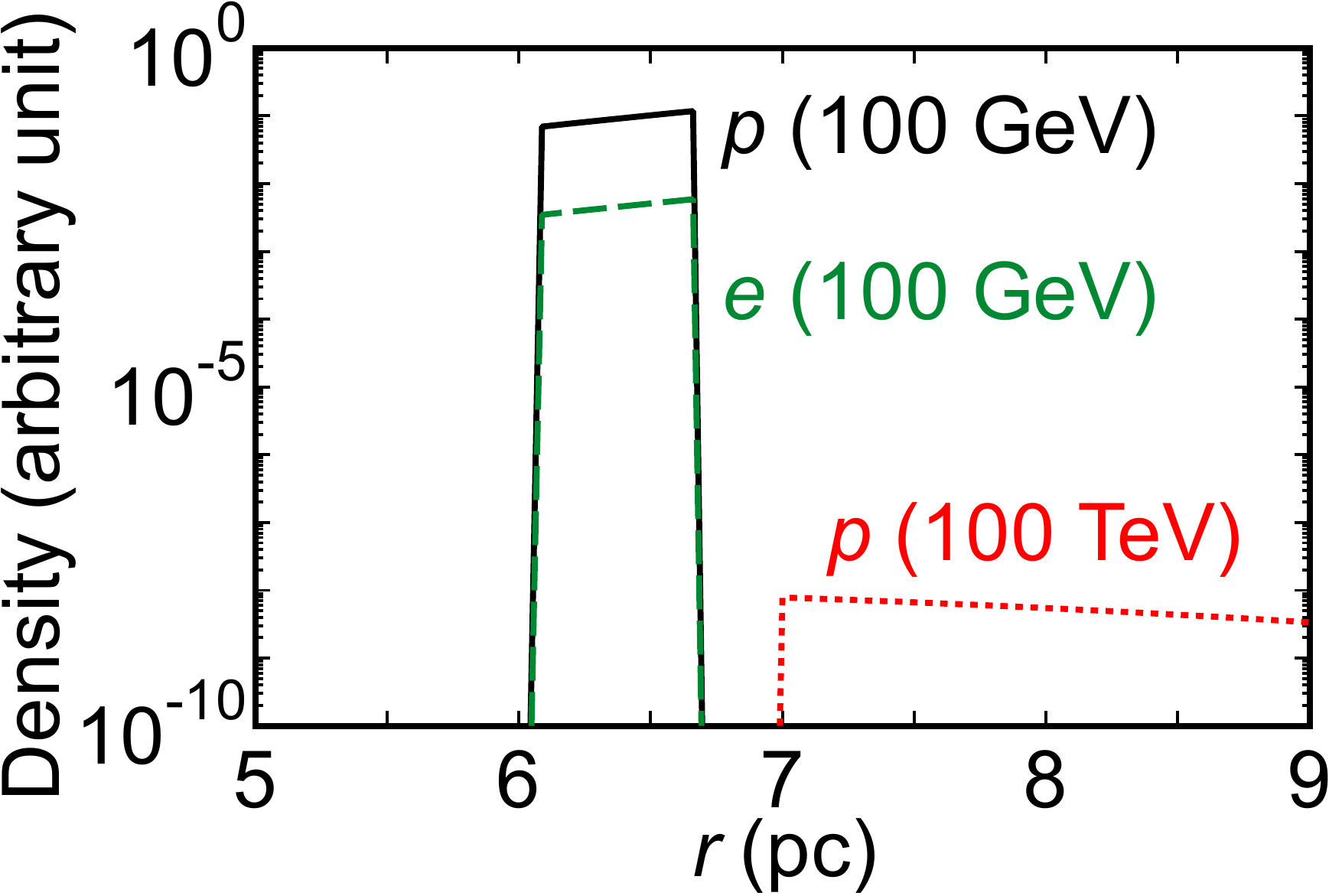} \caption{CR density distributions
at $t=t_{\rm age}$ for CR protons with energy of $E=100$~GeV (solid black
line), $E=100$~TeV (dotted red line), and that for CR electrons
with $E=100$~GeV (dashed green line).}  \label{fig:CRdist}
\end{figure}

\begin{figure}[t]
\includegraphics[width=84mm]{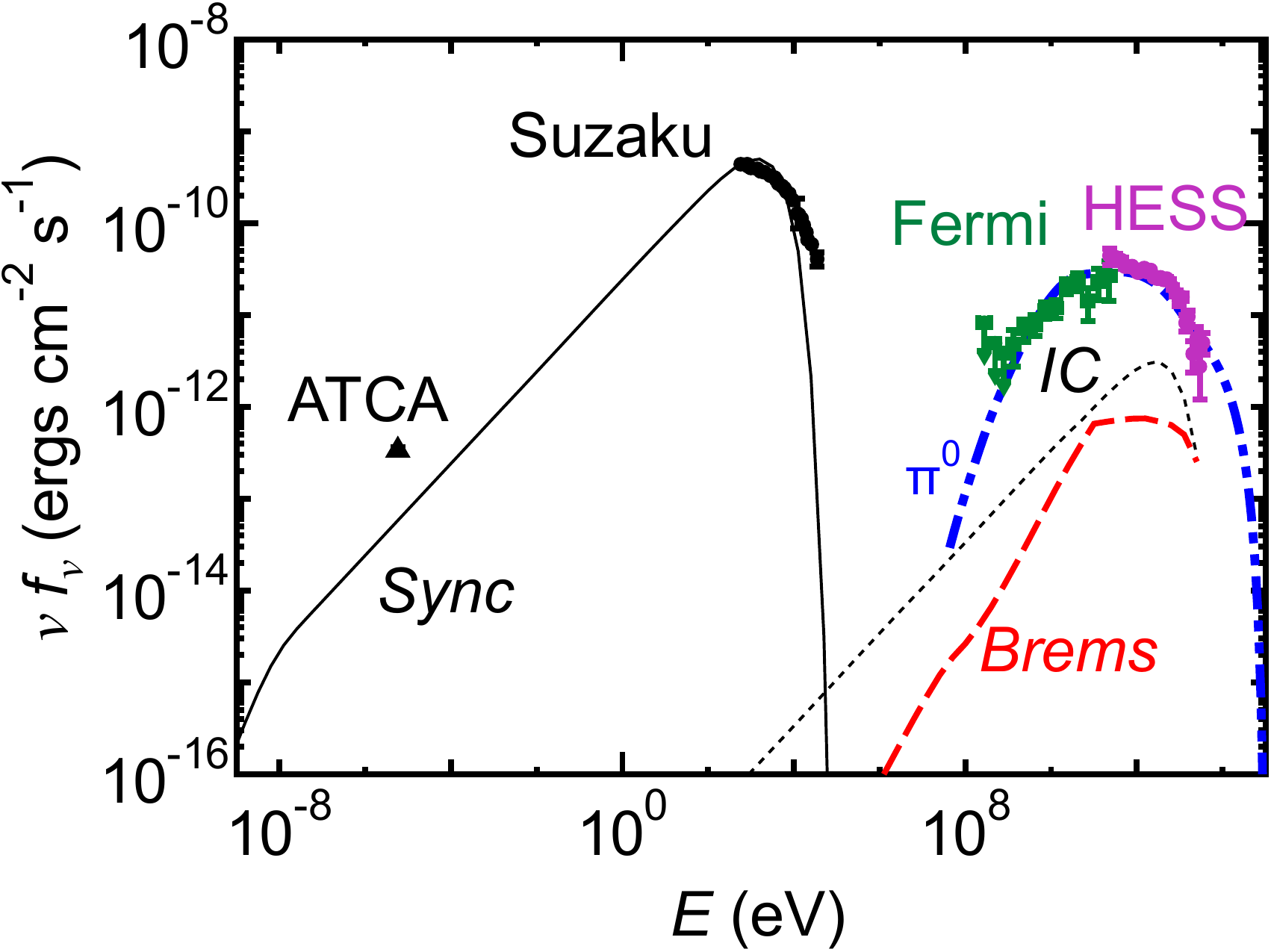} \caption{SED for the fiducial model
SNR. Synchrotron radiation (solid black line), nonthermal
bremsstrahlung (thick dashed red line), and IC scattering (dotted black
line) are from the CR electrons. Hadronic ($\pi^0$ decay) gamma rays are
shown by the two-dots-dashed blue line. Observations of RX~J1713 by
ATCA (black triangle; \citealt{2009A&A...505..157A}), Suzaku (black
circles; \citealt{2008ApJ...685..988T}), Fermi (green squares;
\citealt{2011ApJ...734...28A}), and H.E.S.S. (purple circles;
\citealt{2018A&A...612A...6H}) are also shown.}  \label{fig:SED_fid}
\end{figure}

\begin{figure}[t]
\includegraphics[width=84mm]{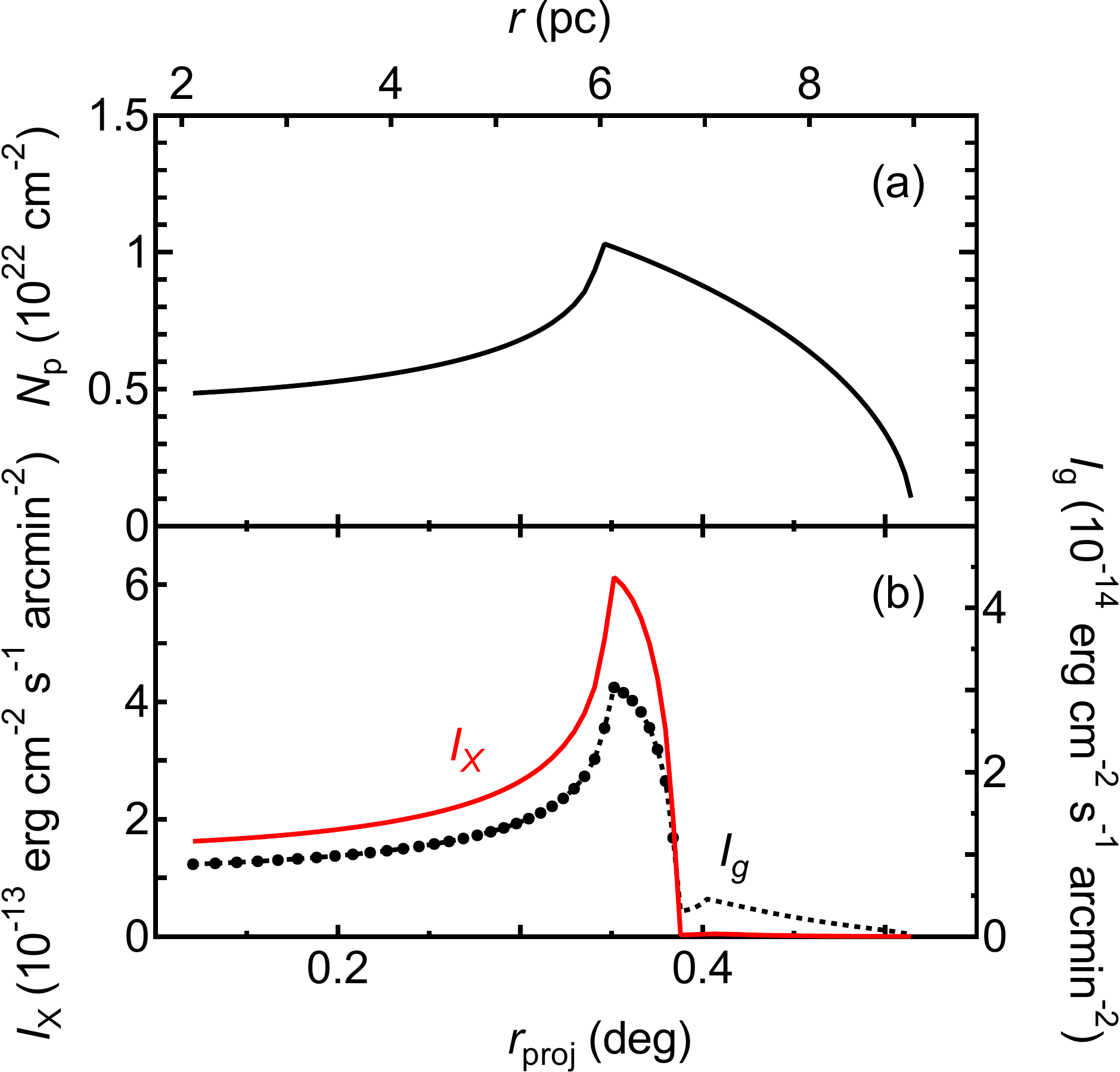} \caption{(a) Profile of the hydrogen
column density $N_p$. (b) Profile of the gamma-ray surface brightness
$I_g$ at 2~TeV (dotted black line) and that of the X-ray surface
brightness $I_X$ at 2~keV (solid red line). Black points on the profile
of $I_g$ show the radial positions of the data in
Figure~\ref{fig:plane_fid}.}  \label{fig:profile}
\end{figure}

In this section, radiations from the model SNR are calculated and they
are compared with observations. We emphasize that we do not aim to
reproduce the observations perfectly, and instead rather attempt to identify
physics behind various observed relations. The shock and escape
radii at present are set at $R_{\rm sh}=6.7$~pc and $R_{\rm
esc}=7.0$~pc, respectively. The current age of the SNR is estimated to
be $t_{\rm age}=1460$~yr (equation~(\ref{eq:Rsh})), and the energy of
CRs currently escaping is $E_{\rm esc}\approx p_{\rm esc}c=34$~TeV
(equation~(\ref{eq:pesc})).  The total energy of CR protons above 1~TeV
is $E_{\rm tot,CR}=5.80\times 10^{49}(250\rm\: cm^{-3}/1\rm\:
cm^{-3})^{-1}$~erg \citep{2018A&A...612A...6H}.

Using the models in Section~\ref{sec:CRdist}, we calculate the CR distributions
and show them in Figure~\ref{fig:CRdist}. CRs with energies of $E<E_{\rm
esc}$ are distributed at the downstream side of the shock. They have not
been swept far downstream and are confined in the ISM ($R_1\lesssim
r<R_{\rm sh}$) because the SNR is relatively young. On the other hand,
CRs with $E>E_{\rm esc}$ have escaped from the SNR and are distributed at
$r>R_{\rm esc}$. However, CR electrons with $E\gg E_{\rm esc}$ have
cooled down mainly because of synchrotron radiation.

Figure~\ref{fig:SED_fid} shows the SED of the model SNR compared with
observations of RX~J1713. Here, we set a threshold energy at $E_{\rm
th}=0.1$~TeV, and assume that only CR protons with energies of $E>E_{\rm
th}$ can enter the clumps. This is because numerical simulations showed
that magnetic fields and/or plasma waves develop around clumps, and 
they prevent lower-energy CRs from intruding into the clumps
\citep{2012ApJ...744...71I,2019ApJ...872...46I}. This means that at a
given radius $r$, CR protons with $E>E_{\rm th}$ interact with the ISM
with an average density $n_{\rm av}$, while those with $E<E_{\rm th}$
remain in the interclump medium with a density $n_{\rm ic}$. We ignore
the volume of the clumps because the filling factor is $x\ll 1$.  On the
other hand, CR electrons may behave differently because amplified
magnetic fields around clumps may cool them down and prevent them from
entering the clumps. In fact, observations showed that synchrotron
emissions originated from electrons decline in large clumps in contrast
with gamma-ray emissions (\citealt{2013ApJ...778...59S}; Sano et al. in
preparation). Thus, we assume that while electrons with $E>E_{\rm th}$
can enter the clumps, 90\% of them have cooled down and cannot
radiate. The following results do not much change as long as the
fraction is large. Most electrons with $E>E_{\rm th}$ remain in and
interact with the interclump medium. Electrons with $E<E_{\rm th}$
cannot enter the clumps and stay in the interclump space. Here, we
implicitly assume that only a ignorable fraction of electrons have
cooled down in the magnetic fields covering the clumps because of their
small volume.

Although we chose $E_{\rm th}=0.1$~TeV to reproduce the
observed SED, it is consistent with the results of numerical simulations
\citep{2019ApJ...872...46I}. This suggests that the
Fe~{\footnotesize I}~K$\alpha$ line emission at 6.40~keV is unlikely to
be observed from the clumps inside the SNR, because while the line
emission is produced through the interaction of MeV CR protons with the
clump gas \citep{2018ApJ...854...87N,2019PASJ...71...78M}, these
low-energy CRs are not allowed to enter the clumps.\footnote{
However, Fe~\footnotesize{I}~K$\alpha$ line emission could be detected
for clumps outside the SNRs because the magnetic fields are not
amplified around the clumps \citep{2021ApJ...908..136F}.}

We calculate radiative processes for electrons using the models by
\citet{2008MNRAS.384.1119F}, and we derive gamma-ray spectra using the
models by \citet{2006ApJ...647..692K}, \citet{2006PhRvD..74c4018K}, and
\citet{2008ApJ...674..278K}. The cosmic microwave background radiation
and a far-infrared component with a temperature $T = 26.5$~K and a
density of $0.415\rm\: eV\: cm^{-3}$ are the seed photon fields for the
IC emission. The latter values are obtained from GALPROP by
\citet{2011ApJ...727...38S} at a distance of 1~kpc. The emissions from
secondary electrons can be ignored.

Figure~\ref{fig:SED_fid} shows that the fiducial model reproduces the
overall trend of the observed SED. While most of the gamma-ray flux is
generated through the hadronic process ($\pi^0$ decay), a small fraction
is attributed to the radiation through the IC scattering. The
contribution of the nonthermal bremsstrahlung to the gamma-ray emission
is negligible. The hadronic gamma-ray flux at $\sim$GeV is lower than
that at $\sim$TeV, which is consistent with the results of the Fermi
observations \citep{2011ApJ...734...28A}. The suppression is attributed
to our assumption that only CR protons with $E>E_{\rm th}\:(=0.1\:\rm
TeV)$ interact with the dense clump medium \citep[see
also][]{2010ApJ...708..965Z,2012ApJ...744...71I,2014MNRAS.445L..70G}.
Although thermal X-ray emissions from RX~J1713 are very dim
\citep{2015ApJ...814...29K}, they are not inconsistent with our model. This
is because the density of the hot interclump medium is low and
because the temperature of the clumps does not increase strongly because the
shocks that propagate in them are weak \citep{2012ApJ...744...71I}.

In Figure~\ref{fig:SED_fid}, the synchrotron flux we predict is
lower than that of the Suzaku observations at $E\gtrsim 10$~keV. This may be
caused by the oversimplification of our model. For example, all CRs with
energies of $>E_{\rm esc}=34$~TeV have escaped from the SNR, while those
with $<34$~TeV are confined within the SNR in our model. If we make a
smoother transition, that is, if we allow some of CRs with $>34$~TeV to
be inside the SNR, the discrepancy disappears. NuSTAR observations may
be useful to discuss this issue \citep[e.g.][]{2019ApJ...877...96T}.  In
order to reproduce both radio and X-ray synchrotron flux observations,
the energy spectrum of electrons rather than the magnetic field strength
needs to be modified. However, the elections responsible for the radio
emission do not contribute to X-rays and gamma rays that are our focus
in this paper. Thus, we neither tune the spectrum nor add another
component for these electrons.

Figure~\ref{fig:profile}(a) shows the profile of hydrogen column density
$N_p$ derived from $n_{\rm av}$. To estimate the projected radius
$r_{\rm proj}$, the distance to the SNR (RX~J1713) is taken to be 1~kpc
\citep{2003PASJ...55L..61F,2004A&A...427..199C,2005ApJ...631..947M}.
The profile of $N_p$ roughly reproduces observations
\citep{2021ApJ...915...84F}. In Figure~\ref{fig:profile}(b), we present
the profile of the gamma-ray surface brightness $I_g$ at 2~TeV and that
of the X-ray surface brightness $I_X$ at 2~keV. While they are generally
similar, the gamma-ray emission is noticed even at $r_{\rm proj}\gtrsim
0.4$~deg, which is not observed for the X-rays. The former is produced
by CR protons escaped from the SNR ($r>R_{\rm sh}$ and $E>E_{\rm
esc}$). Since the magnetic field strength at the upstream side is weaker
than that at the downside ($B_u < B_d$), CR electrons cannot effectively
create synchrotron X-ray emission at $r>R_{\rm sh}$
\citep{2017JHEAp..13...17O}. This is qualitatively consistent with the
results of the H.E.S.S. observations that showed gamma-ray emissions
extending beyond the X-ray-emitting shell \citep{2018A&A...612A...6H}.

\subsection{A Plane Formed in $N_p$--$I_X$--$I_g$ Space}

\begin{figure*}[t]
\plotone{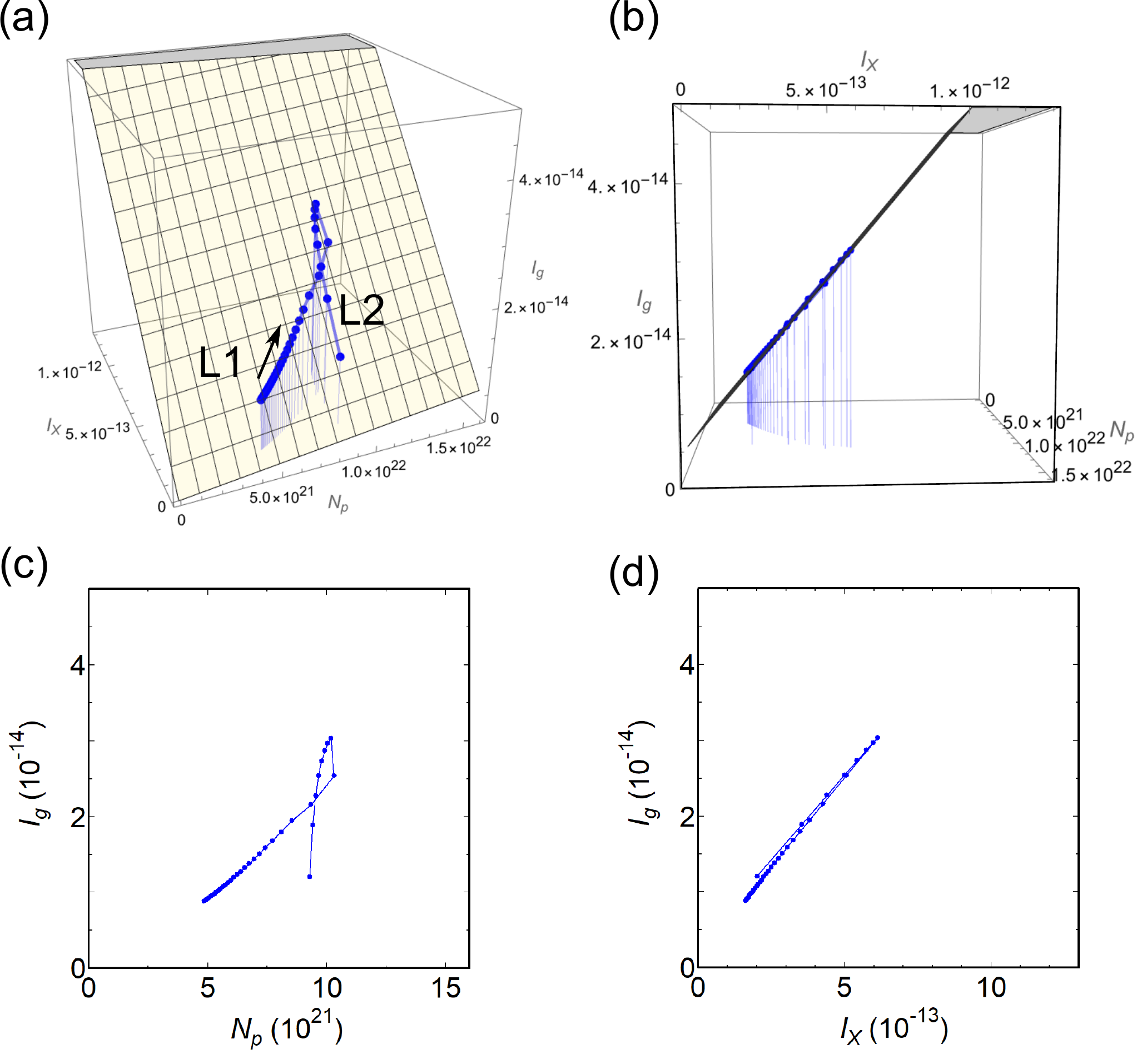} \caption{Points are the predicted $(N_p, I_X,
I_g)$ distributions for the fiducial model with $n_{\rm av}=250\:\rm cm^{-3}$, $K_{eq}=0.05$, and $B_u=12\:\mu\rm G$. (a) 3D view. The
best-fitting plane is also shown. The arrow shows the directions in
which $r_{\rm proj}$ increases. (b) Different view of panel (a). The near side is
indicated by the bold frame. (c) Projection onto the $N_p$--$I_g$
plane. (d) Projection onto the $I_X$--$I_g$ plane. The units of the axes
are $\rm cm^{-2}$ for $N_p$ and $\rm erg\: cm^{-2}\: s^{-1}\: arcmin^{-2}$
for $I_X$ and $I_g$.}  \label{fig:plane_fid}
\end{figure*}

\begin{figure}[t]
\includegraphics[width=84mm]{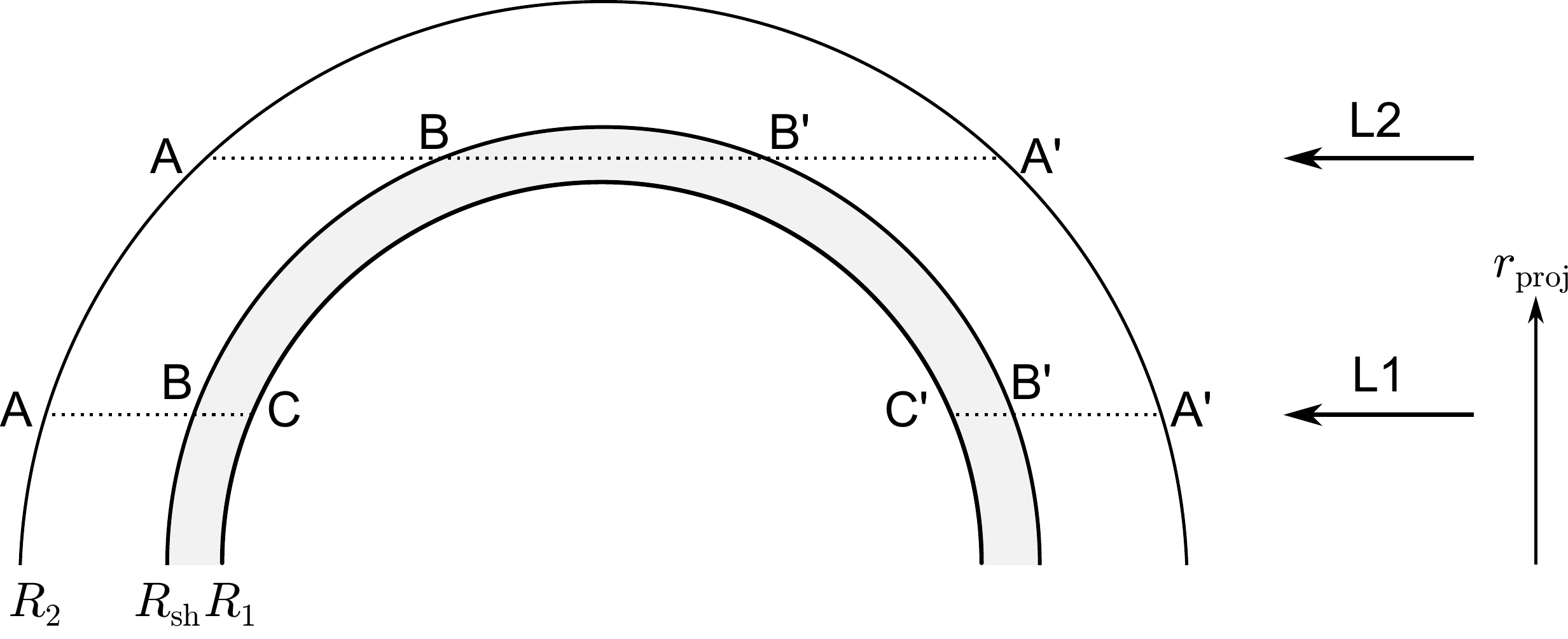} \caption{Cross-section of the
SNR. Unescaped CRs are permeated at $R_1\lesssim r < R_{\rm sh}$, while
the ISM is distributed at $R_1<r<R_2$. For the line of sight L1,
$\tilde{N}_p$ is the column density along the segments BC$+$C'B'
($=\ell$), while $N_p$ is that along AC$+$C'A'. For the line of sight
L2, $\tilde{N}_p$ is the column density along the segment BB'
($=\ell$), while $N_p$ is that along AA'. }  \label{fig:Np}
\end{figure}

\begin{figure*}[t]
\plotone{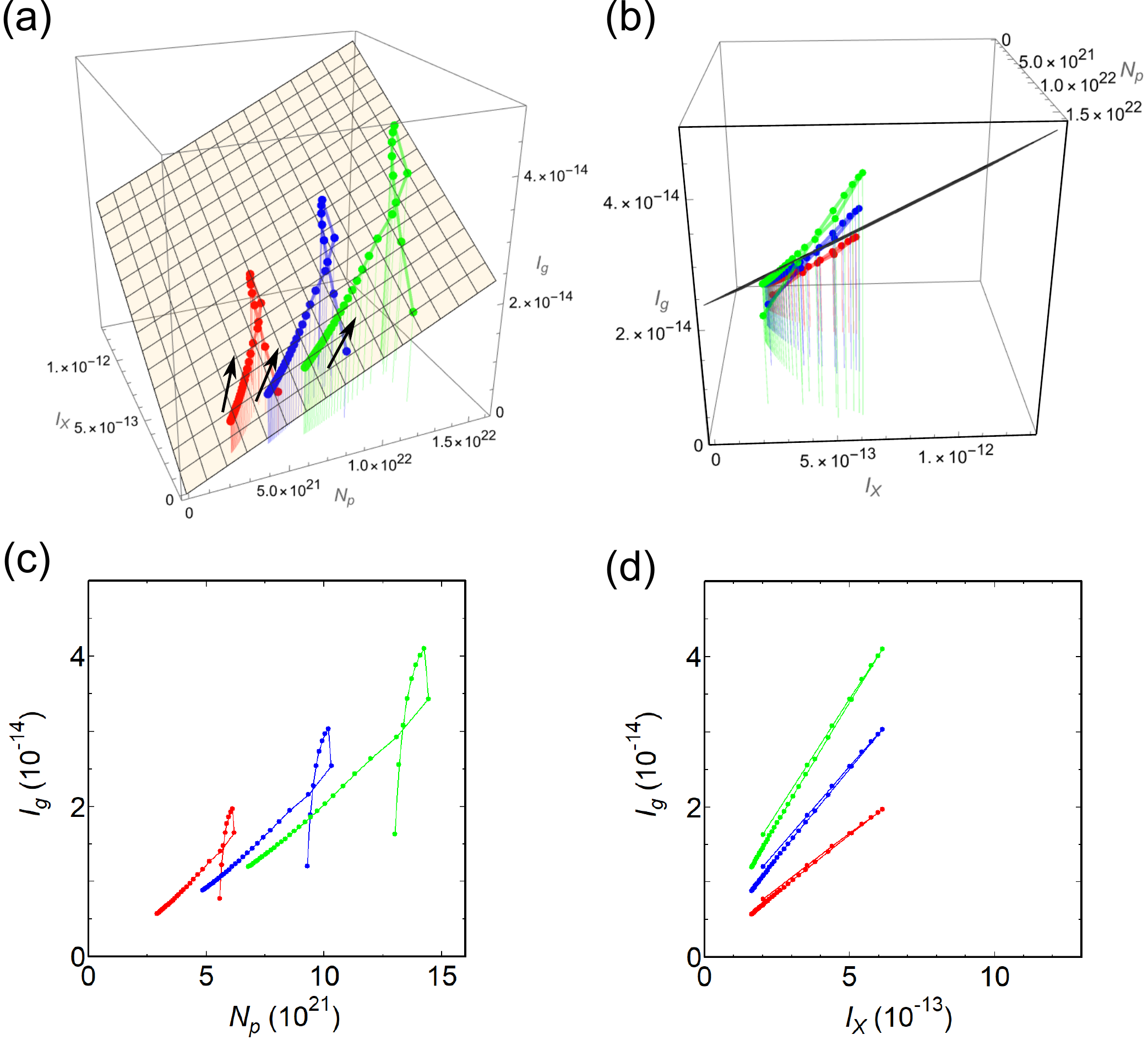} \caption{Same as Figure~\ref{fig:plane_fid} but the
results for low-density ($n_{\rm av}=150\rm\: cm^{-3}$; red) and
high-density ($n_{\rm av}=350\rm\: cm^{-3}$; green) models are added to
be compared with those for the fiducial model ($n_{\rm av}=250\rm\:
cm^{-3}$; blue). The values of $K_{\rm ep}$ and $B_u$ are fixed at the
fiducial values ($K_{\rm ep}=0.05$ and $B_u=12\:\mu\rm G$).}  \label{fig:plane_nav}
\end{figure*}

\begin{figure*}[t]
\plotone{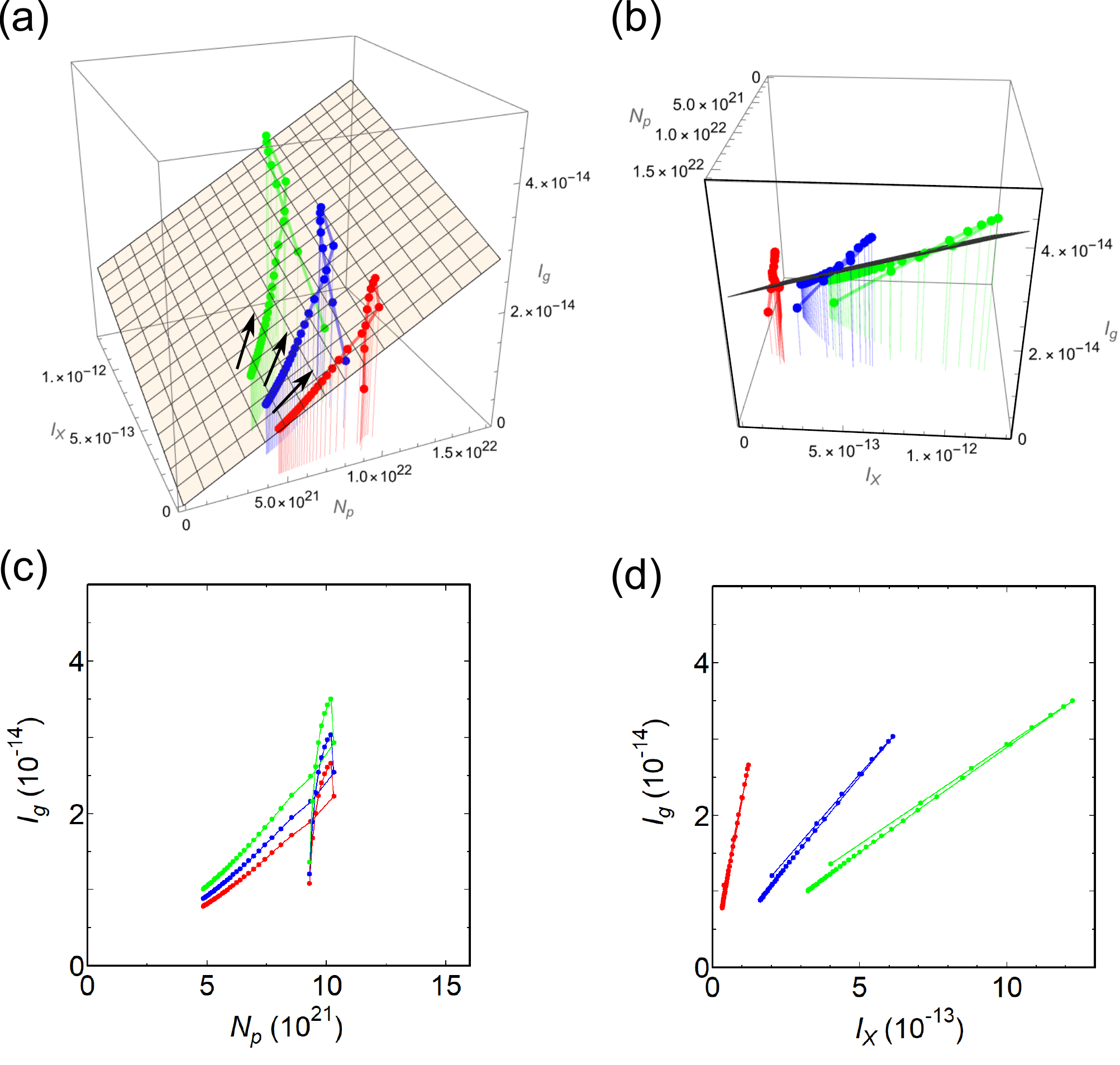} \caption{Same as Figure~\ref{fig:plane_fid} but in
addition to the results for the fiducial model ($K_{\rm ep}=0.05$; blue),
those for low-$K_{\rm ep}$ ($K_{\rm ep}=0.01$; red) and high-$K_{\rm ep}$
($K_{\rm ep}=0.1$; green) models are shown. The values of $n_{\rm av}$ and
$B_u$ are fixed at the fiducial values ($n_{\rm av}=250\rm\: cm^{-3}$ and $B_u=12\:\mu\rm G$).}  \label{fig:plane_Kep}
\end{figure*}

\begin{figure*}[t]
\plotone{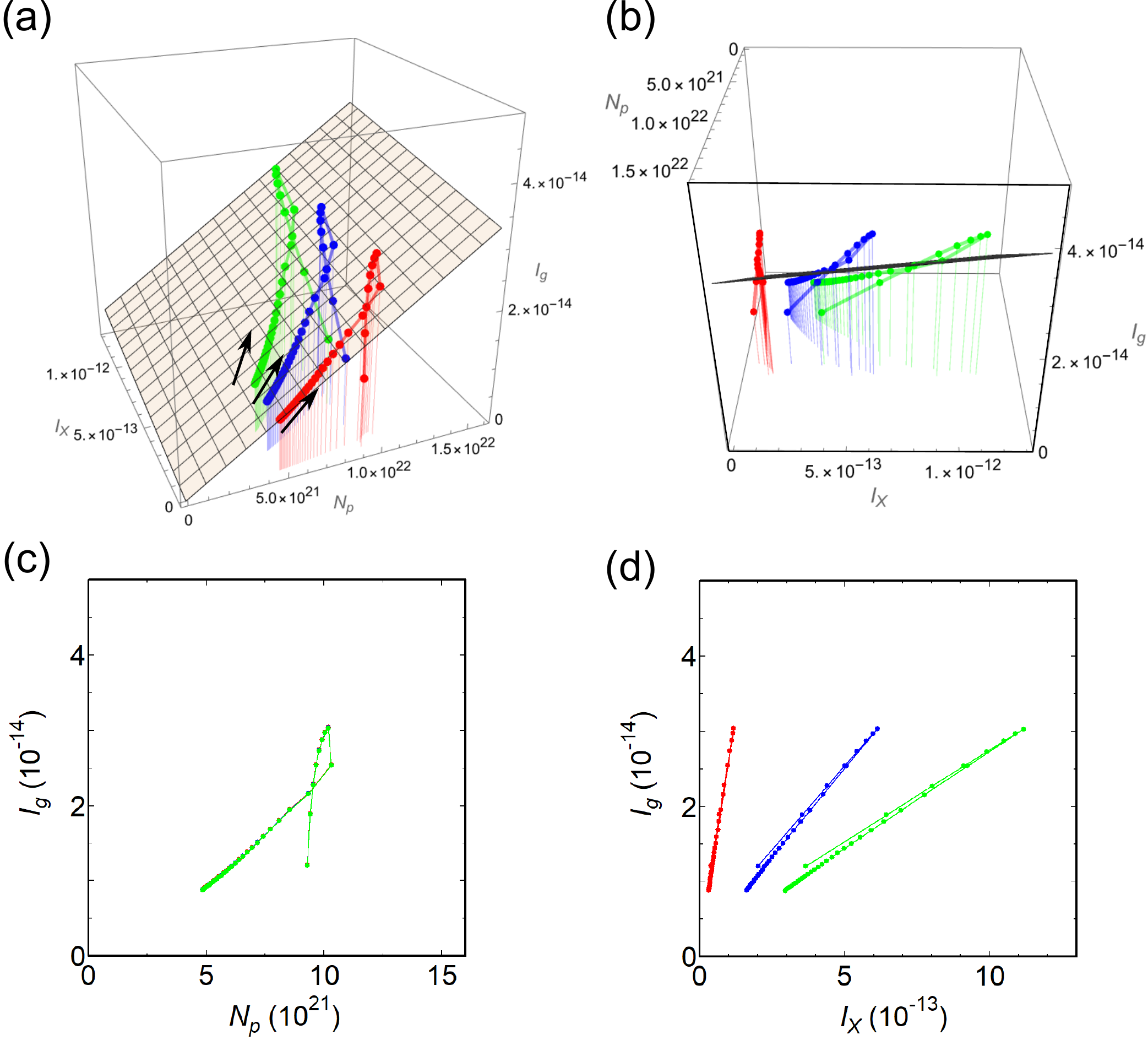} \caption{Same as Figure~\ref{fig:plane_fid} but in
addition to the results for the fiducial model ($B_u=12\:\mu\rm G$;
blue), those for low-$B$ ($B_u=6\:\mu\rm G$; red) and high-$B$
($B_u=16\:\mu\rm G$; green) models are shown. The values of $n_{\rm av}$
and $K_{\rm ep}$ are fixed at the fiducial values ($n_{\rm av}=250\rm\:
cm^{-3}$ and $K_{\rm ep}=0.05$). In panel (c), the data for all models 
overlap.}  \label{fig:plane_B}
\end{figure*}

Recently, \citet{2021ApJ...915...84F} indicated that the pixel data
of RX~J1713 form a plane in the 3D space of the hydrogen column density
($N_p$), the X-ray surface brightness ($I_X$), and the gamma-ray surface
brightness ($I_g$).\footnote{ \citet{2021ApJ...915...84F} used the X-ray
photon counts ($N_X$) and gamma-ray photon counts ($N_g$) to discuss the
plane. However, these observables depend on the observation times. Thus,
we adopt $I_X$ and $I_g$ instead.} They indicated that the plane is
formed if both hadronic and leptonic IC components contribute to the
observed gamma rays. In this section, we discuss the plane in detail
using our model SNR.

In Figures~\ref{fig:plane_fid} we plot $(N_p, I_X, I_g)$ along the
radial direction for the fiducial-model SNR; $I_X$ and $I_g$ are the
values at 2~keV and 2~TeV, respectively. The projected radius ($r_{\rm
proj}$) increases in the directions of the arrow, and $r_{\rm proj}$ of
the individual points, which we call the ``data'', corresponds to those
of the black dots in Figure~\ref{fig:profile}(b). We use the data only
for $r<R_{\rm sh}$ because $I_X$ and $I_g$ are very small outside the
shock (Figure~\ref{fig:profile}(b)) and would be discarded in actual
observations. However, this does not mean that our calculations
for CRs at $r>R_{\rm sh}$ are useless because we have explicitly
confirmed that their contribution can be ignored, even to the projected
emissions from the inside of the SNR ($r_{\rm proj}<R_{\rm sh}$).

Because Figure \ref{fig:plane_fid}(a) shows the results for the fiducial
model, the data are the same as those in
Figure~\ref{fig:profile}. Following \citet{2021ApJ...915...84F}, we fit
the following plane to the data:
\begin{align}
\label{eq:plane1}
& \left(\frac{I_g}{10^{-14}\:\rm erg\:
cm^{-2}\: s^{-1}\: arcmin^{-2}}\right) \hspace{10mm}\nonumber \\
&= a \left(\frac{N_p}{10^{21}\:\rm cm^{-2}}\right) \nonumber \\
&~~~ + b \left(\frac{I_X}{10^{-13}\:\rm erg\:
cm^{-2}\: s^{-1}\: arcmin^{-2}}\right)
\end{align}
where $a$ and $b$ are parameters and the best-fitting plane is shown in
Figure~\ref{fig:plane_fid}(a). The best-fitting parameters are
$a=0.0341$ and $b=0.439$. Figure~\ref{fig:plane_fid}(b) is a different
view of Figure~\ref{fig:plane_fid}(a) and shows that the plane nicely
fits to the data. The standard deviation in the direction of $I_g$ from
the plane is $1.1\times 10^{-16}\rm\: erg\: cm^{-2}\: s^{-1}\:
arcmin^{-2}$. However, the plane normal is almost on the $I_x$--$I_g$
plane in Figure~\ref{fig:plane_fid}, which contradicts Figure~4 in
\citet{2021ApJ...915...84F}. This can be explained as follows. The
hadronic gamma-ray brightness is written $I_{\rm gp}\propto \tilde{N}_p
n_p$, where $\tilde{N}_p$ is the column density of the ISM hydrogens
interacting with CRs. This is because while the ISM is distributed at
$R_1 < r < R_2$, unescaped CRs are permeated only at $R_1\lesssim r <
R_{\rm sh}$ (Figures~\ref{fig:CRdist} and~\ref{fig:Np}). We emphasize
that $\tilde{N}_p < N_p$ and their relation is nonlinear (not
$\tilde{N}_p \propto N_p$); the difference between $\tilde{N}_p$ and
$N_p$ was not considered in \citet{2021ApJ...915...84F}. The leptonic
gamma-ray brightness of bremsstrahlung origin and that of IC origin are
represented by $I_{\rm ge,br}\propto \tilde{N}_p n_e$ and
$I_{\rm ge,IC}\propto n_{\rm ph} n_e \ell$, respectively, where $n_{\rm ph}$ is the
seed photon field and $\ell$ is the depth of the region containing CRs
along the line of sight (Figure~\ref{fig:Np}). On the other hand, because
the brightness of X-ray synchrotron emission is given by
\begin{equation}
\label{eq:Ix}
 I_X\propto B^2 n_e \ell\:,
\end{equation}
we obtain $I_{\rm ge,br}\propto \tilde{N}_p I_X/(B^2 \ell)$ and
$I_{\rm ge,IC} \propto n_{\rm ph} I_X/B^2$. Thus, the gamma-ray
brightness is represented as
\begin{eqnarray}
\label{eq:Ig1}
I_g &=& I_{\rm gp} + I_{\rm ge,IC} + I_{\rm ge,br} \nonumber\\
&=&  f n_p \tilde{N}_p + (g n_{\rm
ph} B^{-2}) I_X \nonumber\\
& &  + (h B^{-2}\ell^{-1}) \tilde{N}_p I_X\:,
\end{eqnarray}
where $f$, $g$, and $h$ are constants. The first, second, and third
terms on the right-hand side represent the hadronic, IC, and
bremsstrahlung components, respectively.

For the fiducial model, Figure~\ref{fig:SED_fid} indicates that the
hadronic component is the main contributor to the gamma-ray
emission. Thus, the gamma-ray brightness is approximated by $I_g \sim
I_{\rm gp} \propto n_p \tilde{N}_p$. Moreover, because we have assumed that
$n_p =K_{\rm ep}^{-1} n_e$ (Section~\ref{sec:inside}), the
gamma-ray brightness should be
\begin{eqnarray}
\label{eq:Ig2}
 I_g &\propto& n_p \tilde{N}_p 
\propto   K_{\rm ep}^{-1} n_e \tilde{N}_p \nonumber\\
 &\propto& I_X \tilde{N}_p/(K_{\rm ep} B^2\ell)
\propto K_{\rm ep}^{-1} n_{\rm av} B^{-2} I_X\:,
\end{eqnarray}
where we used relation~(\ref{eq:Ix}) and $n_{\rm
av}=\tilde{N}_p/\ell$. We emphasize that this relation represents the
hadronic component, although the last one is represented by $I_X$ and
apparently does not include $\tilde{N}_p$. Because $n_{\rm av}=\rm const$ and
$B=\rm const$ in the region in which most CRs exist ($R_1\lesssim r\lesssim
R_{\rm sh}$), we finally obtain the relation of $I_g\propto I_X$, which
explains the similarity between $I_g$ and $I_X$ in
Figure~\ref{fig:profile}(b). In Figure~\ref{fig:plane_fid}(a), the
linear sequence L1 corresponds to the data around the line of sight L1
in Figure~\ref{fig:Np} and $r_{\rm proj}\sim 0.^\circ 1$--$0.^\circ 3$ in
Figure~\ref{fig:profile}. Along this sequence, all of $N_p$, $I_X$ and
$I_g$ gradually increase as $r_{\rm proj}$ increases. On the other hand,
the linear sequence L2 in Figure~\ref{fig:plane_fid}(a) corresponds to
the data around the line of sight L2 in Figure~\ref{fig:Np} and $r_{\rm
proj}\sim 0.^\circ 35$--$0.^\circ 37$ in Figure~\ref{fig:profile}. Along this
sequence, while $N_p$ changes little, $I_X$ and $I_g$ rapidly decreases
as $r_{\rm proj}$ increases. These two sequences form the plane in
Figures~\ref{fig:plane_fid}(a) and (b).

Relation~(\ref{eq:Ig2}) suggests that the $I_X$--$I_g$ relation depends
on $f_{Xg}\equiv K_{\rm ep}^{-1} n_{\rm av} B^{-2}$. This means that if
$f_{Xg}$ varies, the plane shown in Figures~\ref{fig:plane_fid}(a) and
(b) ($I_g\propto f_{Xg} I_X$) should rotate around the $N_p$-axis. This
happens when the SNR and $f_{Xg}$ are not azimuthally uniform (i.e. not
uniform around the SNR center), for example. First, we vary $n_{\rm av}$
while keeping $K_{\rm ep}$ and $B$ unchanged. This modifies $N_p$ because it
depends on $n_{\rm av}$. Specifically, we calculate the emissions from
the SNR when $n_{\rm av}=150\rm\: cm^{-3}$ (low-density model) and
$n_{\rm av}=350\rm\: cm^{-3}$ (high-density model); the other parameters are
the same as those in the fiducial model ($n_{\rm av}=250\rm\:
cm^{-3}$). In Figure~\ref{fig:plane_nav}, we show the data for the
fiducial model and the high- and low-density models.  As $n_{\rm av}$ changes, the
sequence shifts to the direction of $N_p$ and $I_g$ (red and green
sequences in Figures~\ref{fig:plane_nav}(a) and (c)). We fit
equation~(\ref{eq:plane1}) to all the data, and the resultant plane is
shown in Figures~\ref{fig:plane_nav}(a) and (b). The standard deviation
of the data in the direction of $I_g$ from the plane is $2.4\times
10^{-15}\rm\: erg\: cm^{-2}\: s^{-1}\: arcmin^{-2}$. The best-fitting
parameters are $a=0.147$ and $b=0.192$. The plane normal has a
substantial $N_p$-component, which is consistent with Figure~4 in
\citet{2021ApJ...915...84F}. Figure~\ref{fig:plane_nav}(d) clearly shows
that the $I_g$--$I_X$ relation rotates ($I_g\propto f_{Xg} I_X$) as
$n_{\rm av}$ (i.e. $f_{Xg}$) changes. The shift and rotation of the
sequence is the reason of the plane inclination, and they have nothing
to do with CR electrons. Thus, the plane configuration discovered by
\citet{2021ApJ...915...84F} may simply reflect that the ISM around
RX~J1713 is not uniform in the azimuthal direction. In fact,
\citet{2012ApJ...746...82F} found that the ISM has both atomic and
molecular hydrogens and that $N_p$ in the northwest region is larger
than in the southeast region. Therefore, the observed plane could
be reproduced even if the gamma-ray emission from RX~J1713 is purely
hadronic and if $n_{\rm av}$ changes in the azimuthal direction \citep[see also][]{2020MNRAS.499.1154H}. We
note that we do not perform quantitative comparison of the plane angle
with that obtained by \citet{2021ApJ...915...84F}. This is because the
plane fitting to the data is not necessarily good
(Figure~\ref{fig:plane_nav}(b)), and thus the angle depends on the bias
of the chosen data pointa. The SED of the whole SNR can be reproduced if the
average of $n_{\rm av}$ across the SNR is $\sim 250\rm\: cm^{-3}$.

Second, we vary $K_{\rm ep}$ while keeping $n_{\rm av}$ and $B$
unchanged. Specifically, we calculate the emissions from the SNR when
$K_{\rm ep}=0.01$ (low-$K_{\rm ep}$ model) and $K_{\rm ep}=0.1$ (high-$K_{\rm ep}$
model); the other parameters are the same as those in the fiducial model
($K_{\rm ep}=0.05$). The results are shown in
Figure~\ref{fig:plane_Kep}. This time, the sequence shifts mainly in the
direction of $I_X$ (Figures~\ref{fig:plane_Kep}(a) and (d)) because the CR
electron density $n_e$ depends on $K_{\rm ep}$. The sequence also shifts in
the direction of $I_g$ (Figure~\ref{fig:plane_Kep}(c)) because the
contribution of the leptonic IC component changes in the gamma-ray
band. We fit equation~(\ref{eq:plane1}) to all the data, and the
resultant plane is shown in Figures~\ref{fig:plane_Kep}(a) and (b). The
standard deviation of the data in the direction of $I_g$ from the plane
is $2.5\times 10^{-15}\rm\: erg\: cm^{-2}\: s^{-1}\: arcmin^{-2}$. The
best-fitting parameters are $a=0.177$ and $b=0.108$. The plane normal
has a substantial $N_p$-component. Figure~\ref{fig:plane_Kep}(d) shows
that the $I_g$--$I_X$ relation rotates as $K_{\rm ep}$ changes. Again, the
shift and rotation of the sequence is the reason for the plane
inclination. If the plane angle for an SNR is determined by the
variation of $K_{\rm ep}$ as it is in Figure~\ref{fig:plane_Kep}, it could be
used to estimate the average respective contributions of the hadronic
and leptonic components to the total gamma rays, as was done by
\citet{2021ApJ...915...84F}.

Third, we vary $B$ while keeping $n_{\rm av}$ and $K_{\rm ep}$
unchanged. Specifically, we calculate the emissions from the SNR when
the upstream magnetic fields are $B_u=6\:\mu\rm G$ (low-$B$ model) and
$B_u=16\:\mu\rm G$ (high-$B$ model); the other parameters are the same as
those in the fiducial model ($B_u=12\:\mu\rm G$). The results are shown in
Figure~\ref{fig:plane_B}. This time, the sequence shifts only in the
direction of $I_X$ (Figures~\ref{fig:plane_B}(a) and (d)) because the
synchrotron radiation depends on $B$, while $N_p$ and $I_g$ do not
depend on $B$. We fit equation~(\ref{eq:plane1}) to all the data, and the
resultant plane is shown in Figures~\ref{fig:plane_B}(a) and (b). The
standard deviation of the data in the direction of $I_g$ from the plane
is $3.0\times 10^{-15}\rm\: erg\: cm^{-2}\: s^{-1}\: arcmin^{-2}$. The
best-fitting parameters are $a=0.206$ and $b=0.0423$. The plane normal
has a much smaller $I_X$-component than in Figure~4 in
\citet{2021ApJ...915...84F}. This may indicate that the magnetic field
variation is not the cause of the plane inclination for
RX~J1713. Figure~\ref{fig:plane_B}(d) shows that the $I_g$--$I_X$
relation for the fiducial model rotates as $B$ (i.e. $f_{Xg}$) changes.

Figures~\ref{fig:plane_nav}--\ref{fig:plane_B} indicate that the
rotation and the direction of the shift of the radial sequence (red,
blue, and green) could be used to identify the parameter that azimuthally
varies in the SNR, if individual radial sequences are similar. For
example, the sequence shifts both in the $N_p$ and $I_g$ directions when
$n_{\rm av}$ changes (Figures~\ref{fig:plane_nav}(c)), while it does not
shift in the $N_p$ direction when $K_{\rm ep}$ or $B$ change
(Figures~\ref{fig:plane_Kep}(c) and \ref{fig:plane_B}(c)). If individual
radial sequences are not similar, it means that the SNR has a complex
structure. Although it may not be easy, we could still explore the
structure by solving an inversion problem. For example, we could
constrain the structure from the observed pixel data of $(N_p, I_X,
I_g)$ by comparing them with a number of simulation data using deep
learning.

Finally, we note that if the gamma-ray emission from the SNR is
genuinely produced by IC scattering, it should be $I_g\approx I_{\rm ge,
IC} \propto n_{\rm ph} I_X/B^2$, which does not depend on $N_p$
explicitly. If $B=\rm const$ and $I_X$ is irrelevant to $N_p$, we obtain a
plane $I_g\propto I_X$, which is inconsistent with the direction of the
observed plane \citep{2021ApJ...915...84F}. However, it could be 
interesting future work to investigate the possibility that $B$ and/or
$n_e$ have a special dependence on $N_p$ or $n_{\rm av}$ so that the
observed plane is reproduced.

\section{Conclusions}
\label{sec:conc}

We have studied nonthermal emissions from an SNR interacting with
inhomogeneous ISM taking RX~J1713 as an example. Based on an analytical
model, we obtained the distribution of CRs around the SNR. We assumed
that the ISM is composed of high-density clumps immersed in low-density
interclump medium, as is suggested by numerical simulations. Moreover, we
assumed that only high-energy ($>0.1$~TeV) CRs can enter the clumps
because magnetic fields are amplified around the clumps. While
high-energy protons can radiate gamma rays through $pp$-interaction,
most of the high-energy electrons have cooled down because of the amplified
magnetic fields around the clumps, and they do not radiate gamma
rays. Lower-energy CRs cannot penetrate the clumps and remain in the
interclump medium. Our findings are summarized as follows:

\begin{enumerate}
 \item Our model can broadly reproduce the SED of RX~J1713. In
       particular, the observed gamma-ray peak at $\sim $TeV can be
       explained by the exclusion of the lower-energy protons from the
       clumps.

 \item If the SNR is spherically symmetric and if the hadronic component
       is dominant in the gamma rays, the radial sequence of pixel data
       forms a plane in the 3D space formed by the ISM column density,
       the X-ray surface brightness, and the gamma-ray surface
       brightness. However, the plane angle is inconsistent with that
       observationally obtained by \citet{2021ApJ...915...84F}.

 \item If the ISM density, the electron-to-proton ratio, or the magnetic
       field strength is not spherically uniform in the SNR, the plane
       angle significantly changes from the one at which the SNR is
       spherically uniform. In particular, if the ISM density or the
       electron-to-proton ratio is not uniform, the plane angle is
       qualitatively consistent with observed angles. By studying the
       rotation and the shift of the radial sequence of data in the 3D
       space, we could identify which parameter is not uniform.
\end{enumerate}

In the era of the Cherenkov Telescope Array \citep[CTA;][]{2013APh....43....3A},
the intrusion of CRs into the clumps and the structure of the plane will
be studied in more detail for even more SNRs because gamma-ray
observations with higher angular resolutions become available
\citep{2017ApJ...840...74A}.

\begin{acknowledgments}
We would like to thank the anonymous referee for a constructive
report. We thank Y. Fukui and H. Sano for useful discussion. This work
was supported by JSPS KAKENHI No.18K03647, 20H00181, 22H00158, 22H01268
(Y.F.), 22H01251 (R.Y.), and 19H01893, 21H04487 (Y.O.).
\end{acknowledgments}

\bibliography{plane}{}
\bibliographystyle{aasjournal}

\end{document}